\documentclass[twocolumn,prc,showpacs,preprintnumbers,amsmath,amssymb,superscriptaddress,nofootinbib]{revtex4-2}

\usepackage{graphicx}
\usepackage{dcolumn}
\usepackage{bm}
\usepackage{booktabs}
\usepackage{placeins}
\usepackage{rotating}
\usepackage{color}

\begin{document}


\title{Investigating nuclear structure near $N = 32$ and $N = 34$: Precision mass measurements of neutron-rich Ca, Ti and V isotopes}
\author{W. S. Porter}
\email[Corresponding author: ]{wporter@triumf.ca}
    \altaffiliation{Current address: Department of Physics and Astronomy, University of Notre Dame, Notre Dame, Indiana 46656, USA}
    \affiliation{TRIUMF, 4004 Wesbrook Mall, Vancouver, British Columbia V6T 2A3, Canada}
    \affiliation{Department of Physics \& Astronomy, University of British Columbia, Vancouver, British Columbia V6T 1Z1, Canada} 

\author{E. Dunling}
    \altaffiliation{Part of doctoral thesis, University of York, published 2021}
    \affiliation{TRIUMF, 4004 Wesbrook Mall, Vancouver, British Columbia V6T 2A3, Canada}
    \affiliation{Department of Physics, University of York, York, YO10 5DD, United Kingdom}

\author{E. Leistenschneider}
    \altaffiliation{Current address: CERN, 1121, Geneva 23, Switzerland}
    \affiliation{Facility for Rare Isotope Beams, Michigan State University, East Lansing, Michigan 48824, USA}
    \affiliation{National Superconducting Cyclotron Laboratory, Michigan State University, East Lansing, Michigan 48824, USA}

\author{J. Bergmann}
    \affiliation{II. Physikalisches Institut, Justus-Liebig-Universit\"{a}t, 35392 Gie{\ss}en, Germany}
    
\author{G. Bollen}
    \affiliation{Facility for Rare Isotope Beams, Michigan State University, East Lansing, Michigan 48824, USA}
    \affiliation{National Superconducting Cyclotron Laboratory, Michigan State University, East Lansing, Michigan 48824, USA}
    \affiliation{Department of Physics and Astronomy, Michigan State University, East Lansing, Michigan 48824, USA}

\author{T. Dickel}
    \affiliation{II. Physikalisches Institut, Justus-Liebig-Universit\"{a}t, 35392 Gie{\ss}en, Germany}
    \affiliation{GSI Helmholtzzentrum f\"{u}r Schwerionenforschung GmbH, Planckstra{\ss}e 1, 64291 Darmstadt, Germany}
    
\author{K. A. Dietrich}
    \affiliation{TRIUMF, 4004 Wesbrook Mall, Vancouver, British Columbia V6T 2A3, Canada}
    \affiliation{Ruprecht-Karls-Universit\"{a}t Heidelberg, D-69117 Heidelberg, Germany}

\author{A. Hamaker}
    \affiliation{Facility for Rare Isotope Beams, Michigan State University, East Lansing, Michigan 48824, USA}
    \affiliation{National Superconducting Cyclotron Laboratory, Michigan State University, East Lansing, Michigan 48824, USA}
    \affiliation{Department of Physics and Astronomy, Michigan State University, East Lansing, Michigan 48824, USA}
    
\author{Z. Hockenbery}
    \affiliation{TRIUMF, 4004 Wesbrook Mall, Vancouver, British Columbia V6T 2A3, Canada}
    \affiliation{Department of Physics, McGill University, 3600 Rue University, Montr\'eal, QC H3A 2T8, Canada}
    
\author{C. Izzo}
    \affiliation{TRIUMF, 4004 Wesbrook Mall, Vancouver, British Columbia V6T 2A3, Canada}

\author{A. Jacobs}
    \affiliation{TRIUMF, 4004 Wesbrook Mall, Vancouver, British Columbia V6T 2A3, Canada}
    \affiliation{Department of Physics \& Astronomy, University of British Columbia, Vancouver, British Columbia V6T 1Z1, Canada}
    
\author{A. Javaji}
    \affiliation{TRIUMF, 4004 Wesbrook Mall, Vancouver, British Columbia V6T 2A3, Canada}
    \affiliation{Department of Physics \& Astronomy, University of British Columbia, Vancouver, British Columbia V6T 1Z1, Canada} 
    
\author{B. Kootte}
    \affiliation{TRIUMF, 4004 Wesbrook Mall, Vancouver, British Columbia V6T 2A3, Canada}
    \affiliation{Department of Physics \& Astronomy, University of Manitoba, Winnipeg, Manitoba R3T 2N2, Canada}
    
\author{Y. Lan}
    \affiliation{TRIUMF, 4004 Wesbrook Mall, Vancouver, British Columbia V6T 2A3, Canada}
    \affiliation{Department of Physics \& Astronomy, University of British Columbia, Vancouver, British Columbia V6T 1Z1, Canada}
   
\author{I. Miskun}
    \affiliation{II. Physikalisches Institut, Justus-Liebig-Universit\"{a}t, 35392 Gie{\ss}en, Germany}
    
\author{I. Mukul}
    \affiliation{TRIUMF, 4004 Wesbrook Mall, Vancouver, British Columbia V6T 2A3, Canada}
    
\author{T. Murb\"{o}ck}
    \affiliation{TRIUMF, 4004 Wesbrook Mall, Vancouver, British Columbia V6T 2A3, Canada}

\author{S. F. Paul}
    \affiliation{TRIUMF, 4004 Wesbrook Mall, Vancouver, British Columbia V6T 2A3, Canada}
    \affiliation{Ruprecht-Karls-Universit\"{a}t Heidelberg, D-69117 Heidelberg, Germany}

\author{W. R. Pla\ss}
    \affiliation{II. Physikalisches Institut, Justus-Liebig-Universit\"{a}t, 35392 Gie{\ss}en, Germany}
    \affiliation{GSI Helmholtzzentrum f\"{u}r Schwerionenforschung GmbH, Planckstra{\ss}e 1, 64291 Darmstadt, Germany}
    
\author{D. Puentes}
    \affiliation{Facility for Rare Isotope Beams, Michigan State University, East Lansing, Michigan 48824, USA}
    \affiliation{National Superconducting Cyclotron Laboratory, Michigan State University, East Lansing, Michigan 48824, USA}
    \affiliation{Department of Physics and Astronomy, Michigan State University, East Lansing, Michigan 48824, USA}

\author{M. Redshaw}
    \affiliation{National Superconducting Cyclotron Laboratory, Michigan State University, East Lansing, Michigan 48824, USA}
    \affiliation{Department of Physics, Central Michigan University, Mount Pleasant, Michigan 48859, USA}
    
\author{M. P. Reiter}
    \affiliation{TRIUMF, 4004 Wesbrook Mall, Vancouver, British Columbia V6T 2A3, Canada}
    \affiliation{II. Physikalisches Institut, Justus-Liebig-Universit\"{a}t, 35392 Gie{\ss}en, Germany}
    \affiliation{School of Physics and Astronomy, University of Edinburgh, Edinburgh, EH9 3FD, United Kingdom}

\author{R. Ringle}
    \affiliation{Facility for Rare Isotope Beams, Michigan State University, East Lansing, Michigan 48824, USA}
    \affiliation{National Superconducting Cyclotron Laboratory, Michigan State University, East Lansing, Michigan 48824, USA}
    
\author{J. Ringuette}
    \affiliation{TRIUMF, 4004 Wesbrook Mall, Vancouver, British Columbia V6T 2A3, Canada}
    \affiliation{Department of Physics, Colorado School of Mines, Golden, Colorado 80401, USA}

\author{R. Sandler}
    \affiliation{Department of Physics, Central Michigan University, Mount Pleasant, Michigan 48859, USA}
    
\author{C. Scheidenberger}
    \affiliation{II. Physikalisches Institut, Justus-Liebig-Universit\"{a}t, 35392 Gie{\ss}en, Germany}
    \affiliation{GSI Helmholtzzentrum f\"{u}r Schwerionenforschung GmbH, Planckstra{\ss}e 1, 64291 Darmstadt, Germany}
    \affiliation{Helmholtz Forschungsakademie Hessen f\"{u}r FAIR (HFHF), GSI Helmholtzzentrum f\"{u}r Schwerionenforschung, Campus Gie{\ss}en, 35392 Gie{\ss}en, Germany}
    
\author{R. Silwal}
    \altaffiliation{Current address: Department of Physics and Astronomy, Appalachian State University, Boone, North Carolina 28608, USA}
    \affiliation{TRIUMF, 4004 Wesbrook Mall, Vancouver, British Columbia V6T 2A3, Canada}

\author{R. Simpson}
    \affiliation{TRIUMF, 4004 Wesbrook Mall, Vancouver, British Columbia V6T 2A3, Canada}
    \affiliation{Department of Physics \& Astronomy, University of British Columbia, Vancouver, British Columbia V6T 1Z1, Canada}

\author{C. S. Sumithrarachchi}
    \affiliation{Facility for Rare Isotope Beams, Michigan State University, East Lansing, Michigan 48824, USA}
    \affiliation{National Superconducting Cyclotron Laboratory, Michigan State University, East Lansing, Michigan 48824, USA}

\author{A. Teigelh\"{o}fer}
    \affiliation{TRIUMF, 4004 Wesbrook Mall, Vancouver, British Columbia V6T 2A3, Canada}
    
\author{A. A. Valverde}
    \affiliation{Department of Physics \& Astronomy, University of Manitoba, Winnipeg, Manitoba R3T 2N2, Canada}
    
\author{R. Weil}
    \affiliation{Department of Physics \& Astronomy, University of British Columbia, Vancouver, British Columbia V6T 1Z1, Canada}
    
\author{I. T. Yandow}
    \affiliation{Facility for Rare Isotope Beams, Michigan State University, East Lansing, Michigan 48824, USA}
    \affiliation{National Superconducting Cyclotron Laboratory, Michigan State University, East Lansing, Michigan 48824, USA}
    \affiliation{Department of Physics and Astronomy, Michigan State University, East Lansing, Michigan 48824, USA}
    
\author{J. Dilling}
    \affiliation{TRIUMF, 4004 Wesbrook Mall, Vancouver, British Columbia V6T 2A3, Canada}
    \affiliation{Department of Physics \& Astronomy, University of British Columbia, Vancouver, British Columbia V6T 1Z1, Canada}
    
\author{A. A. Kwiatkowski}
    \affiliation{TRIUMF, 4004 Wesbrook Mall, Vancouver, British Columbia V6T 2A3, Canada}
    \affiliation{Department of Physics and Astronomy, University of Victoria, Victoria, British Columbia V8P 5C2, Canada}
    

\date{\today}

\begin{abstract}
Nuclear mass measurements of isotopes are key to improving our understanding of nuclear structure across the chart of nuclides, in particular for the determination of the appearance or disappearance of nuclear shell closures. We present high-precision mass measurements of neutron-rich Ca, Ti and V isotopes performed at TRIUMF’s Ion Trap for Atomic and Nuclear science (TITAN) and the Low Energy Beam and Ion Trap (LEBIT) facilities. These measurements were made using the TITAN multiple-reflection time-of-flight mass spectrometer (MR-ToF-MS) and the LEBIT 9.4T Penning trap mass spectrometer. In total, 13 masses were measured, eight of which represent increases in precision over previous measurements. These measurements refine trends in the mass surface around $N = 32$ and $N = 34$, and support the disappearance of the $N = 32$ shell closure with increasing proton number. Additionally, our data does not support the presence of a shell closure at $N = 34$.

\end{abstract}

\maketitle

\section{Introduction}

The structure of nuclei far from stability is of primary interest within low-energy nuclear physics. The Nuclear Shell Model has long been established as the backbone of isotopic structure, correctly reproducing the canonical magic numbers ($N,Z = 2,8,20,28,50,82$) for protons and neutrons for spherical-like nuclei. Outside of these established points of strongest binding, magic behavior has appeared and disappeared at many asymmetric proton-to-neutron ratios \cite{Tanihata1985,Kanungo2013}. The establishment of shell closures across the chart of nuclides is critical in part for the benchmarking of nuclear theories \cite{Georgieva2015,Simonis2016}. 

Previous experimental and theoretical studies have confirmed the presence of a closed shell at $N = 32$ near the proton $Z = 20$ shell. In particular, large 2$^+$ excitation energies \cite{Huck1985, Cortes2020}, small $B(E2; 0^+ \to 2^+)$ \cite{Seidlitz2011,Goldkuhle2019} and nuclear mass trends \cite{Wienholtz2013,Reiter2018,Leistenschneider2021} have all indicated the presence of a subshell at $N = 32$. Nuclear theories have corroborated experimental findings of such a subshell as well \cite{Otsuka2001,Leistenschneider2018,Li2020}.

The existence of a similar subshell at $N = 34$ remains an open question. Experimental evidence from excitation energies \cite{Steppenbeck2013} support the presence of a subshell. However, trends in two-neutron separation energies from mass measurements, most recently of Sc isotopes in \cite{Leistenschneider2021}, generally do not support the existence of such a subshell.  Data from \cite{Michimasa2018} suggests the presence of the $N = 34$ subshell in the Ca isotopes, whereas current Ti and V data do not suggest such a closure in their isotope chains. More precise experimental data is required in the region to establish the existence or non-existence of an $N = 34$ subshell closure.

In this article, we present high-precision nuclear mass measurements of neutron-rich Ca, Ti and V isotopes completed in collaboration at TRIUMF’s Ion Trap for Atomic and Nuclear
science (TITAN) and the Low Energy Beam and Ion Trap (LEBIT) facility at the National Superconducting Cyclotron
Laboratory.

\section{Experiment}

At TRIUMF, mass measurements were performed at TITAN \cite{Dilling2006} using the Multiple-Reflection Time-of-Flight Mass Spectrometer (MR-ToF-MS) \cite{Jesch2015}. Isotopes of interest were produced at TRIUMF's Isotope Separator and Accelerator (ISAC) \cite{Ball2016}, where a 480 MeV, 50 $\mu$A proton beam was impinged on a $22.7$~g/cm$^2$ thick Ta target. Stopped spallation and fragmentation products diffused out of the target towards a hot Re ion source, where they were surface ionized. Further ionization for Ti isotopes was achieved via TRIUMF's resonant ionization laser ion source (TRILIS), via a two-step resonant laser excitation scheme \cite{Lassen2017}. All ionized beams were sent to a mass separator which removed non-isobaric products. The isobaric beam was then transported to the TITAN facility, where it was cooled and bunched via the TITAN radio-frequency quadrupole (RFQ) cooler and buncher, a linear RFQ filled with inert He gas at 10$^{-2}$ mbar \cite{Brunner2012}. Cooled ion bunches were sent to the TITAN MR-ToF-MS for mass measurement.

The TITAN MR-ToF-MS determines the masses of ions via their time-of-flight over a given path and kinetic energy \cite{Wollnik1990,Plab2013}. Since the mass resolution is proportional to the time-of-flight ($R = t/2\Delta t$), a long flight path is desired and achieved via two electrostatic mirrors. Isochronous reflection of ions by these mirrors for a sufficient number of turns achieves the desired resolution \cite{Reiter2021}.

The TITAN MR-ToF-MS consists of two primary sections; a preparation section and an analyzer section. Cooled ion bunches were injected into the preparation section, which consists of a series of RFQs which further cool ion bunches for 3 ms before delivery to the analyzer section. Once inside the mass analyzer, bunches underwent between 350 and 520 isochronous turns between mirrors for a total time-of-flight of $\sim 10$ ms before ejection onto MagneToF detector \cite{Stresau2006} for time-of-flight detection. To minimize contaminant species in our spectra, a mass range selector consisting of two electrodes inside the mass analyzer deflected away any remaining non-isobaric beam products \cite{Dickel2015}. To measure and detect masses of very low signal-to-background ratios ($< 1$ to $ 10^{4}$), mass-selective re-trapping was employed, where ions of interest are dynamically recaptured in the injection trap after a defined flight time in the mass analyzer \cite{Dickel2017}. Recaptured ions are subsequently re-injected into the mass analyzer for a time-of-flight measurement. This process is highly mass-selective, and allows for the separation of isobaric contaminants from ions of interest during the mass measurement process \cite{Dickel2017,Reiter2021,Beck2021,Izzo_In2021}.

\begin{figure}[tb]
    \begin{center}
        \includegraphics[width=0.9\columnwidth]{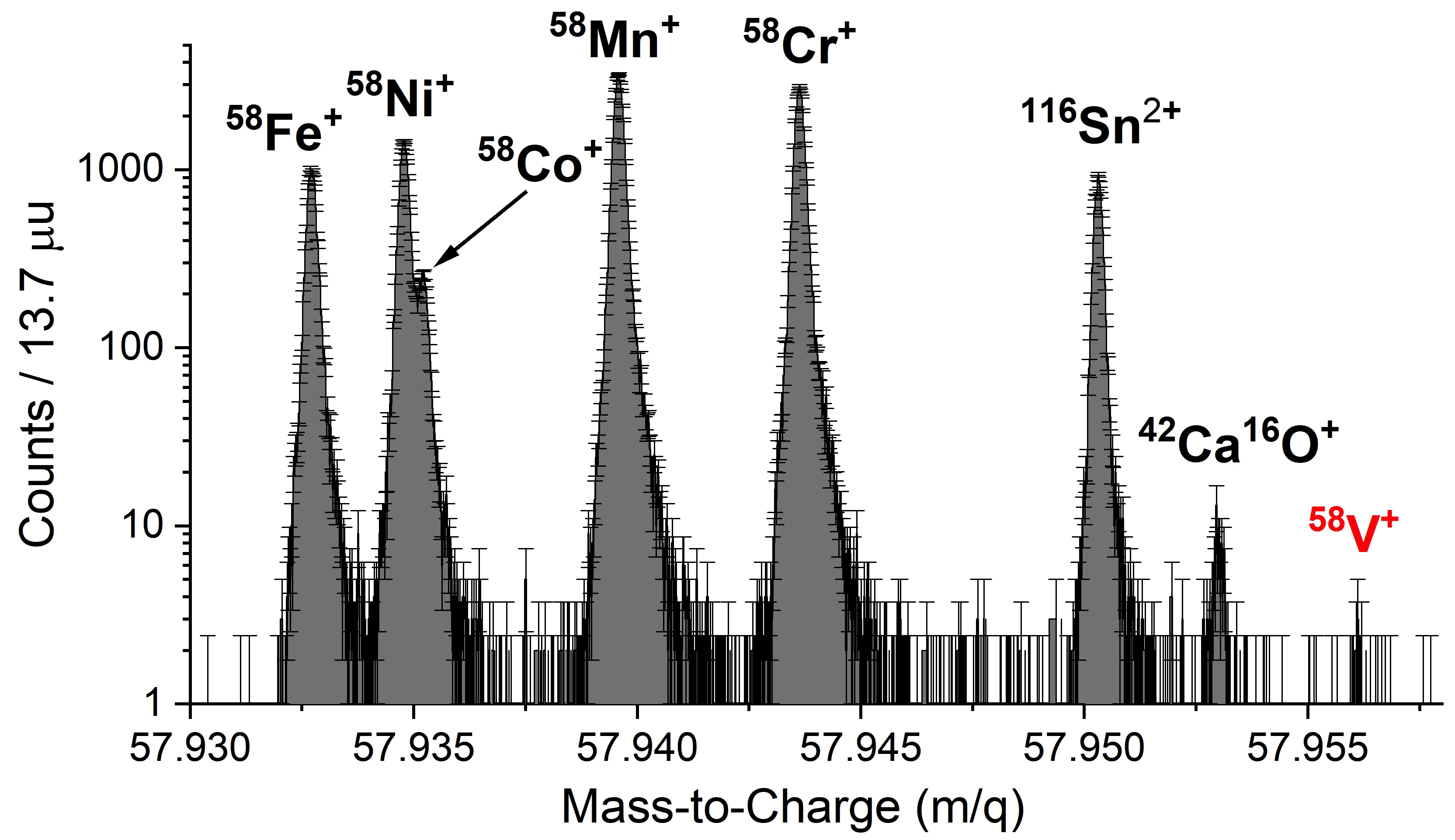}
        \caption{A sample mass spectrum taken at $A = 58$ with the TITAN MR-ToF-MS. The identified ion species are labeled.}
        \label{fig:58u_spectra}
        \includegraphics[width=0.8\columnwidth]{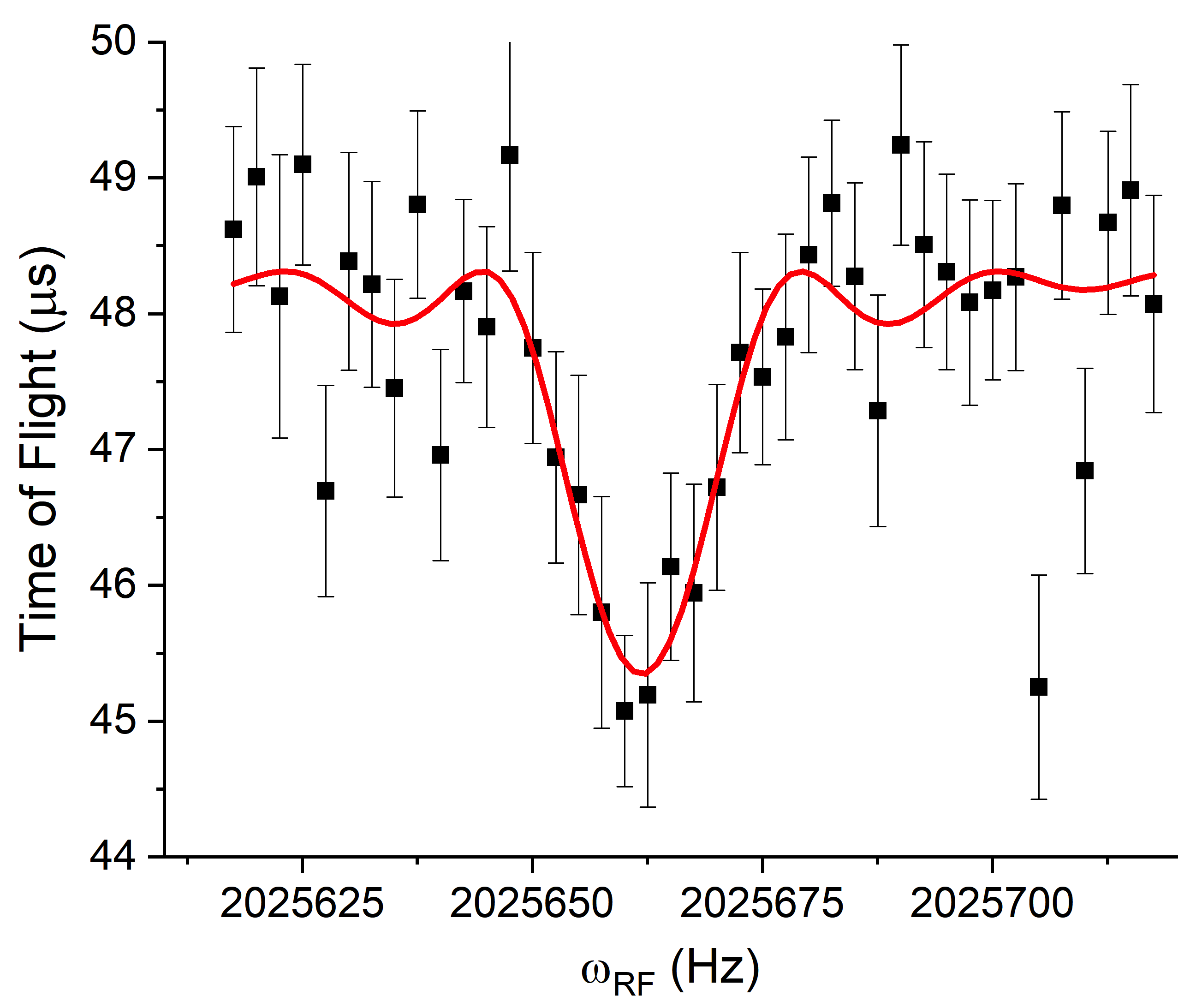}
        \caption{A ToF-ICR spectrum taken for $^{55}$Ti$^{16}$O$^+$ with the LEBIT Penning trap mass spectrometer. The fit result of an analytical function described in \cite{Konig1995}, shown in red, is used to extract the minimum in time-of-flight that occurs at $\omega_{rf} = \omega_c$.}
        \label{fig:tof_icr_curve}
    \end{center}
\end{figure}

Delivered beams contained many atomic and molecular contaminant species alongside the Ca, Ti and V species of interest. An example of a mass spectrum taken during the campaign is shown in Fig.~\ref{fig:58u_spectra}. An initial beam assessment was done during the on-line experiment to identify and assign species to all peaks present in the spectrum. Species identity was confirmed by the occurrence of a species across multiple mass units. The identity of Ti in a spectrum was determined via subsequent measurements with TRILIS lasers off and on. In a lasers off measurement, one of the TRILIS lasers was blocked, removing one of the resonant excitation steps and preventing ionization of Ti. This resulted in a decrease in the overall Ti yield, and subsequently a decrease in time-of-flight counts of Ti in a spectrum. A large count increase ($\sim$ a factor 4) during a lasers on measurement unambiguously confirmed the identification of Ti. More details on the experimental campaign of Ca, Ti and V masses at TITAN can be found in \cite{Dunling2021}.

At the NSCL, isotopes of interest were produced via the in flight method \cite{Aysto2001}, where a 130 MeV/u $^{76}$Ge primary ion beam was impinged on a natural Be target $\sim 0.4$~g/cm$^2$ thick. Desired fragments were separated from contaminants via the A1900 fragment separator \cite{Morrissey2003}, and progressed towards a gas catcher \cite{Sumithrarachchi2020}, where they were stopped as ions in a high-purity He gas. Ions were extracted from the gas catcher as a low-energy, continuous beam and transported to a dipole magnet mass separator for separation of non-isobaric products. Ions of interest, which were primarily singly-charged oxides formed from interactions inside the gas catcher, were sent to the LEBIT facility \cite{Ringle2013}.

Continuous beams entering the LEBIT facility were sent to a cooler and buncher for cooling, accumulation, and bunching, and released as ion bunches \cite{Schwarz2016}. Bunches were delivered to the LEBIT Penning trap mass spectrometer, where isobaric contaminants were cleaned away via application of a dipolar RF field \cite{Kwiatkowski2015} before the mass measurement.

Mass measurements of ions within the LEBIT Penning trap were completed via determination of the cyclotron frequency ($\omega_c$) of the ion around a 9.4 T magnetic field using the time-of-flight ion-cyclotron-resonance (ToF-ICR) technique \cite{Dilling2018}. In ToF-ICR, an ion's slow magnetron drift is converted into a fast modified cyclotron motion through the application of a quadrupole radio-frequency (RF) pulse at $\omega_{rf}$ near the cyclotron frequency. At $\omega_{rf} = \omega_c$, and with a well-chosen duration of the RF pulse, full conversion to the fast modified cyclotron motion can be achieved, resulting in a maximum in radial kinetic energy and a minimum in time-of-flight to a downstream detector. Scanning $\omega_{rf}$ over a series of frequencies around $\omega_c$, results in a time-of-flight spectra with a minimum at $\omega_c$, as seen in Figure \ref{fig:tof_icr_curve}. Standard excitation schemes with RF pulse durations between 50 ms and 500 ms, described in more detail in \cite{Konig1995}, were used.

\section{Analysis}

Time-of-flight spectra taken with the TITAN MR-ToF-MS are converted to mass spectra via the relationship:

\begin{equation}
\frac{m_{\text{ion}}}{q} = C(t_{\text{ion}}-t_0)^2
\end{equation}

where $t_0$ is 0.18 $\mu$s as determined offline and $C$ is a calibration factor determined from a high-statistics, well-known reference ion in each spectrum as listed in Table \ref{tab:mass_table}. To account for time-dependent drifts of times-of-flight, a time-resolved calibration (TRC) was performed \cite{Ayet2019} using the mass data acquisition software package \cite{Dickel2019}. Peak centroids were determined by fitting hyper-EMG functions \cite{Purushothaman2017} using the {\sc emgfit} Python library \cite{emgfit}. Statistical uncertainties are generated based on techniques described in \cite{Ayet2019,emgfit}. Systematic uncertainties of the MR-ToF-MS system are described in detail in \cite{Reiter2021,Ayet2019}, and total to a value of $\sim 2.0 \times10^{-7}$. This uncertainty is dominated by the uncertainties due to ion-ion interactions ($3.3 \times10^{-8}$ per detected ion), the non-ideal switching of mirrors ($ 7.0 \times10^{-8}$), and a further unknown systematic error ($\sim 1.9 \times10^{-7}$).

\begin{figure}[tb]
    \begin{center}
        \includegraphics[width=1\columnwidth]{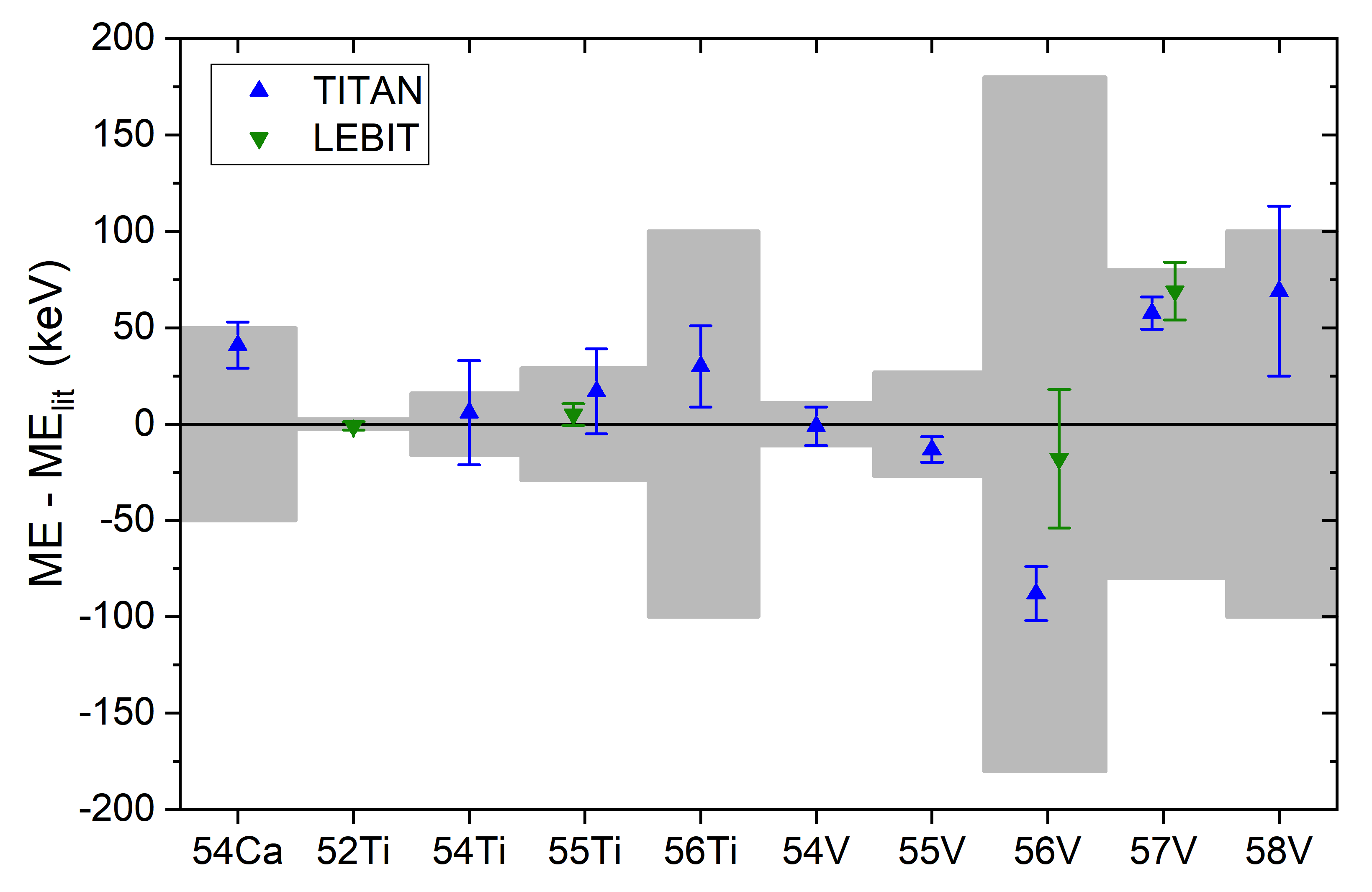}
        \caption{A plot comparing mass values from \cite{Wang2021} to experimental mass results from this work. Grey bands represent error bars given on values in \cite{Wang2021}.}
        \label{fig:mass_compAME}
    \end{center}
\end{figure}

\begin{table*}[ht]

  \centering
  \caption{Results of the mass measurements performed, compared to the values recommended by AME2020 \cite{Wang2021}. Included is the mass ratio ($m_{\text{ionic,IOI}}/m_{\text{ionic,ref}}$) between the ionic masses of the ion of interest (IOI) and the reference ion for all measurements. Differences are $m_{\text{new}} - m_{\text{lit}}$. All mass values are in keV. All TITAN and LEBIT measurements were measured as singly-charged ions. 
  }
    \begin{tabular}{c c c c c c c}
    \toprule    Facility & Nuclide & Mass Excess & Literature & Difference & $\quad$ Reference Ion & Mass Ratio \\    \hline

 TITAN & $^{54}$Ca &  -25119\,\,\,\,\,\,(12) & -25160\,\,\,\,\,\,(50) & 41(51) &  	$\quad$  $^{54}$Cr\,$^{+}$	 	&  1.000\,633\,250\,\,\,\,(239)	  \\
 
 LEBIT & $^{52}$Ti &  -49478.6(2.2) & -49477.7(2.7) & -0.9(3.5) &  	$\quad$  $^{48}$Ti\,$^{16}$O\,$^{+}$     &  0.999\,979\,3643\,\,(887)	  \\
 
 TITAN & $^{54}$Ti &  -45738\,\,\,\,\,\,(27) & -45744\,\,\,\,\,\,(16) & 6(31) &  	$\quad$  $^{54}$Cr\,$^{+}$ 	&  1.000\,222\,872\,\,\,\,(546)     \\
 
 LEBIT & $^{55}$Ti &  -41827.0(5.7) & -41832\,\,\,\,\,\,(29) & 5(30) &  	$\quad$  $^{46}$Ti\,$^{19}$F\,$^{+}$	 	&  0.999\,922\,402\,\,\,\,(133)	  \\
 
 TITAN & $^{55}$Ti &  -41815\,\,\,\,\,\,(22) & -41832\,\,\,\,\,\,(29) & 17(36) &  	$\quad$  $^{55}$Cr\,$^{+}$	 	&  1.000\,259\,788\,\,\,\,(428)	  \\
 
 TITAN & $^{56}$Ti &  -39390\,\,\,\,\,\,(21) & -39420\,\,\,\,(100) & 30(102) &  	$\quad$  $^{56}$Cr\,$^{+}$	 	&  1.000\,305\,043\,\,\,\,(401)	  \\
 
 TITAN & $^{54}$V &  -49899\,\,\,\,\,\,(10) & -49898\,\,\,\,\,\,(11) & -1(15) &  	$\quad$  $^{54}$Cr\,$^{+}$	 	&  1.000\,140\,049\,\,\,\,(195)	  \\

 TITAN & $^{55}$V &  -49138.2(6.6) & -49125\,\,\,\,\,\,(27) & -13(28) &  	$\quad$ $^{55}$Cr\,$^{+}$	&  1.000\,116\,697\,\,\,\,(129)  \\
 
 TITAN & $^{56}$V &  -46268\,\,\,\,\,\,(14) & -46180\,\,\,\,(180) & -88(181) &    $\quad$ $^{56}$Cr\,$^{+}$	&  1.000\,173\,049\,\,\,\,(274)	 \\
  
 LEBIT & $^{56}$V &  -46198\,\,\,\,\,\,(36) & -46180\,\,\,\,(180) & -18(184) &    $\quad$ $^{34}$S\,$^{19}$F$_2$\,$^{+}$	&  0.999\,731\,060\,\,\,\,(540)	 \\
 
 TITAN & $^{57}$V & -44382.4(8.3) & -44440\,\,\,\,\,\,(80) & 58(80) &  $\quad$ $^{57}$Cr\,$^{+}$		&  1.000\,153\,511\,\,\,\,(161)	 \\ 
 
 LEBIT & $^{57}$V & -44371\,\,\,\,\,\,(15) & -44440\,\,\,\,\,\,(80) & 69(81) &  $\quad$ $^{12}$C$_3$\,$^{1}$H$_5$\,$^{16}$O$_2$$^{+}$		&
 0.998\,881\,613\,\,\,\,(220)	 \\ 
 
 TITAN & $^{58}$V & -40361\,\,\,\,\,\,(44) & -40430\,\,\,\,\,\,(100) & 69(109)  &  $\quad$ $^{58}$Fe\,$^{+}$	&  1.000\,403\,867\,\,\,\,(820) \\ 
 \hline

    \end{tabular}%
  \label{tab:mass_table}%
\end{table*}

ToF-ICR spectra taken with the LEBIT Penning trap, e.g. Fig.~\ref{fig:tof_icr_curve}, were fit with the analytical function described in \cite{Konig1995} to extract the cyclotron frequency. The atomic mass is determined via the ratio:

\begin{equation}
R = \frac{\omega_{\text{c,ref}}}{\omega_c} = \frac{m - m_e}{m_{\text{ref}} - m_e}
\end{equation}

where $m_e$ is the mass of the electron, $\omega_c$ and $m$ the cyclotron frequency and mass of the ion of interest, and $\omega_{\text{c,ref}}$ and $m_{\text{ref}}$ the cyclotron frequency and mass of a reference ion. Reference ions were required to have well-known literature masses (with values taken from \cite{Wang2021}) and the same $A/q$ as the ion of interest, and were measured either before or after ion-of-interest measurements to account for potential fluctuations of the magnetic field. Ionization potentials and molecular binding energies are not considered in any of our mass determinations, as they are smaller than 20 eV and do not contribute at the level of precision we attain. Systematic uncertainties, for example, arising from a small misalignment of the trapping axis and the magnetic field, or small deviations from a perfect quadrupole electric potential were determined to contribute at the $\sim 10^{-10}$ level, and thus were negligible \cite{Gulyuz2015}.

\section{Results}

\begin{figure}[tb]
    \begin{center}
        \includegraphics[width=1\columnwidth]{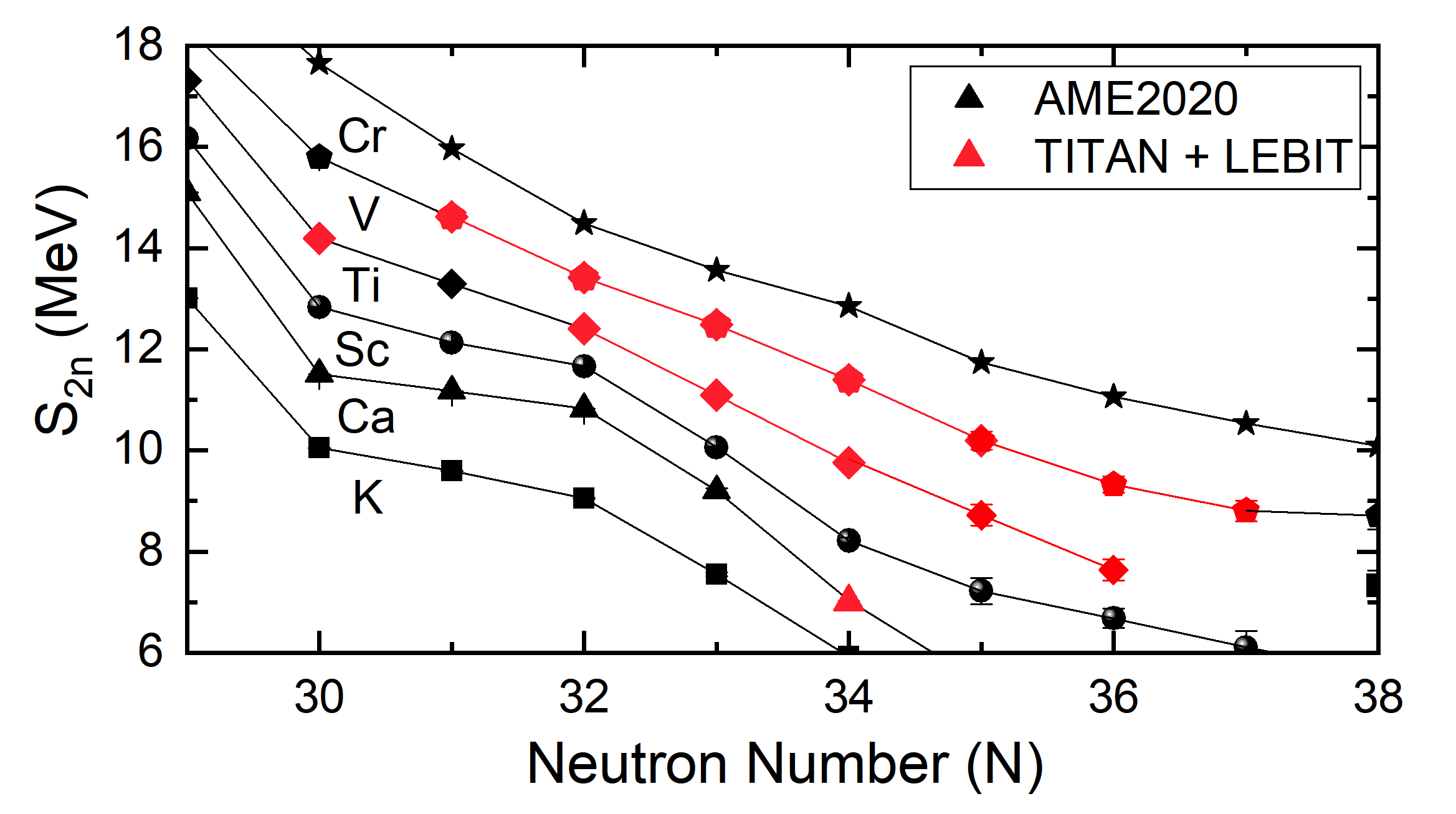}
        \caption{A graph of the $S_{2n}$ in the $Z = 19-24$ isotope chains based on mass values from \cite{Wang2021} (black symbols), and Ca, Ti and V mass measurements from this work (red symbols). For the cases of $^{55}$Ti, $^{56}$V and $^{57}$V where both TITAN and LEBIT measured a mass value, the more precise mass value was used in the determination of the plotted $S_{2n}$ value.}
        \label{fig:s2n}
    \end{center}
\end{figure}

Table \ref{tab:mass_table} reports the masses of all isotopes measured in this work, as well their mass excesses as found in literature. Figure \ref{fig:mass_compAME} compares our mass results to data presented in \cite{Wang2021}. All masses had been previously measured in some capacity to varying precisions; our measurements are in agreement with the earlier results within $0.9\sigma$. Our measurements of $^{54}$Ca, $^{52}$Ti, $^{55}$Ti, $^{56}$Ti, $^{55}$V, $^{56}$V, $^{57}$V and $^{58}$V represent an increase in precision over previous values, with all but two of these representing a precision increase of a factor two or more.

Mass measurements of $^{55}$Ti, $^{56}$V and $^{57}$V were completed at both facilities, with results from both campaigns reported in Table \ref{tab:mass_table}. The reported measurements for $^{55}$Ti, $^{56}$V and $^{57}$V agree to 0.5$\sigma$, 1.8$\sigma$ and 0.7$\sigma$, respectively.

\begin{figure}[tb]
    \begin{center}
        \includegraphics[width=1\columnwidth]{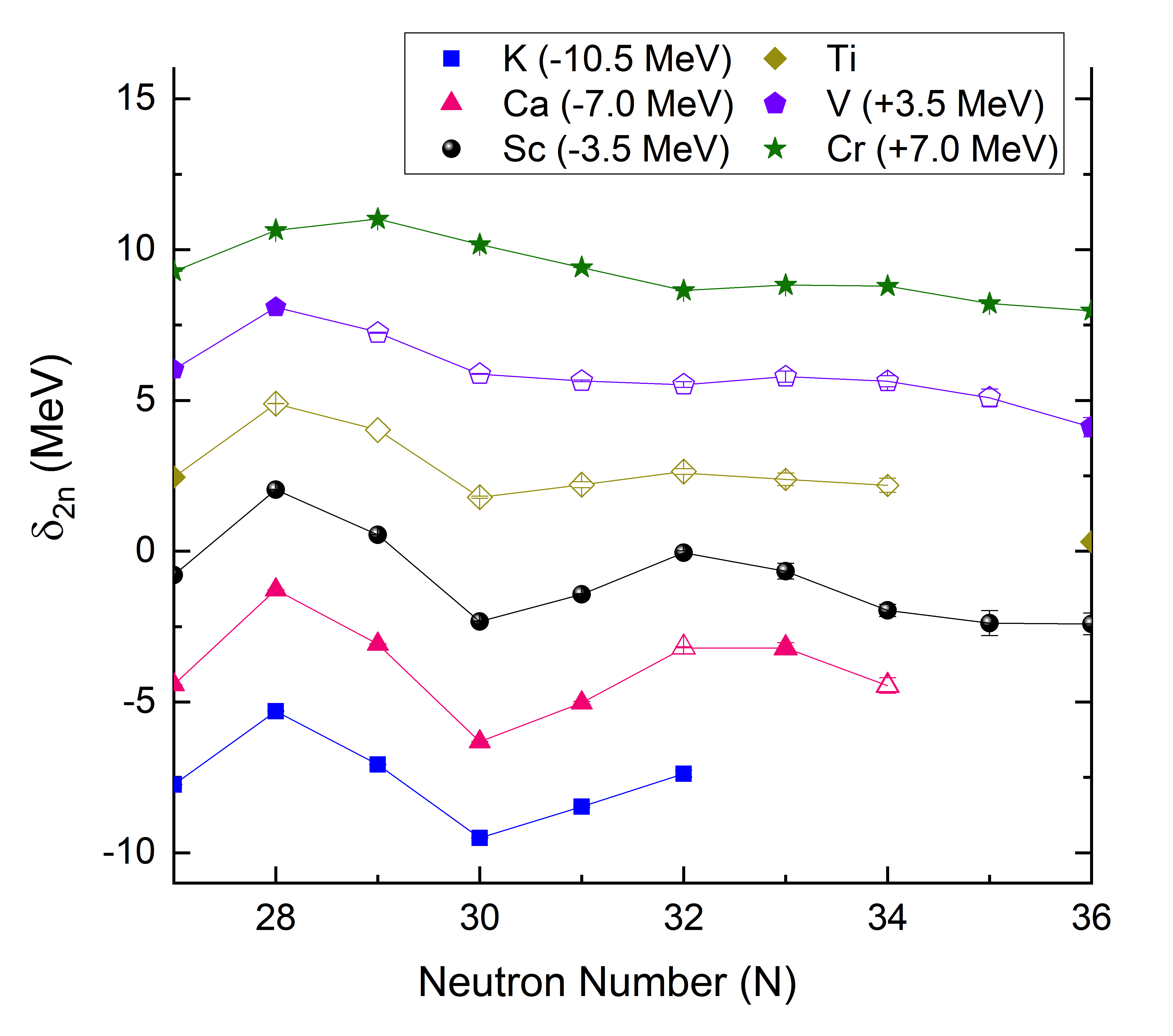}
        \caption{A graph of the $\delta_{2n}$ for the $Z = 19-24$ isotope chains. Values are offset by amounts as presented in the legend. Unfilled symbols represent values from this work. The disappearance of a peak at $N = 32$, signifying the disappearance of a shell closure at $N = 32$, can be seen as proton number increases.}
        \label{fig:delta2n}
    \end{center}
\end{figure}

In order to probe the shell structure in the region encompassed by our measurements, we consider the two-neutron separation energies ($S_{2n}$), defined as:

\begin{equation}
S_{2n}(N,Z) = m(N-2,Z) + 2m_n - m(N,Z)
\end{equation}

with $m(N,Z)$ the mass excess and $m_n$ the mass of the neutron. Figure \ref{fig:s2n} shows $S_{2n}$ values for the $Z = 19-24$ isotopes. Our results show that the steep drop-off at $N = 32$ seen in the Ca and Sc isotopes chains is not present in the Ti and V chains. This confirms the disappearance of the $N = 32$ shell closure at and above $Z = 22$ as presented in \cite{Leistenschneider2018}. No drop-off is seen at $N = 34$ for any of the presented isotope chains.

To further probe the structures seen in separation energy trends, we consider the empirical shell gap parameter ($\delta_{2n}$), given as:

\begin{equation}
    \delta_{2n} = S_{2n}(N,Z) - S_{2n}(N+2,Z)
\end{equation}

 Figure \ref{fig:delta2n} shows $\delta_{2n}$ for the $Z = 19-24$ isotopes. As a pseudo-derivative of $S_{2n}$, low $\delta_{2n}$ values decreasing towards zero indicate a flattening in $S_{2n}$, whereas a sharp peak indicates a steep drop-off in $S_{2n}$ at $N$. Such sharp peaks, when at even neutron numbers, are typically indicative of shell closures \cite{Leistenschneider2018}. A decrease in $\delta_{2n}$ from Ca towards V is seen at $N = 32$, strongly signaling the disappearance of the $N = 32$ shell closure. Additionally, our data does not show any peak-like behavior at $N = 34$, and thus does not support the presence of a shell closure at $N = 34$. However, mass measurements of more neutron-rich isotopes are needed to extend the mass surface and fully characterize the region surrounding $N = 34$.

\begin{figure}[tb]
    \begin{center}
        \includegraphics[width=0.84\columnwidth]{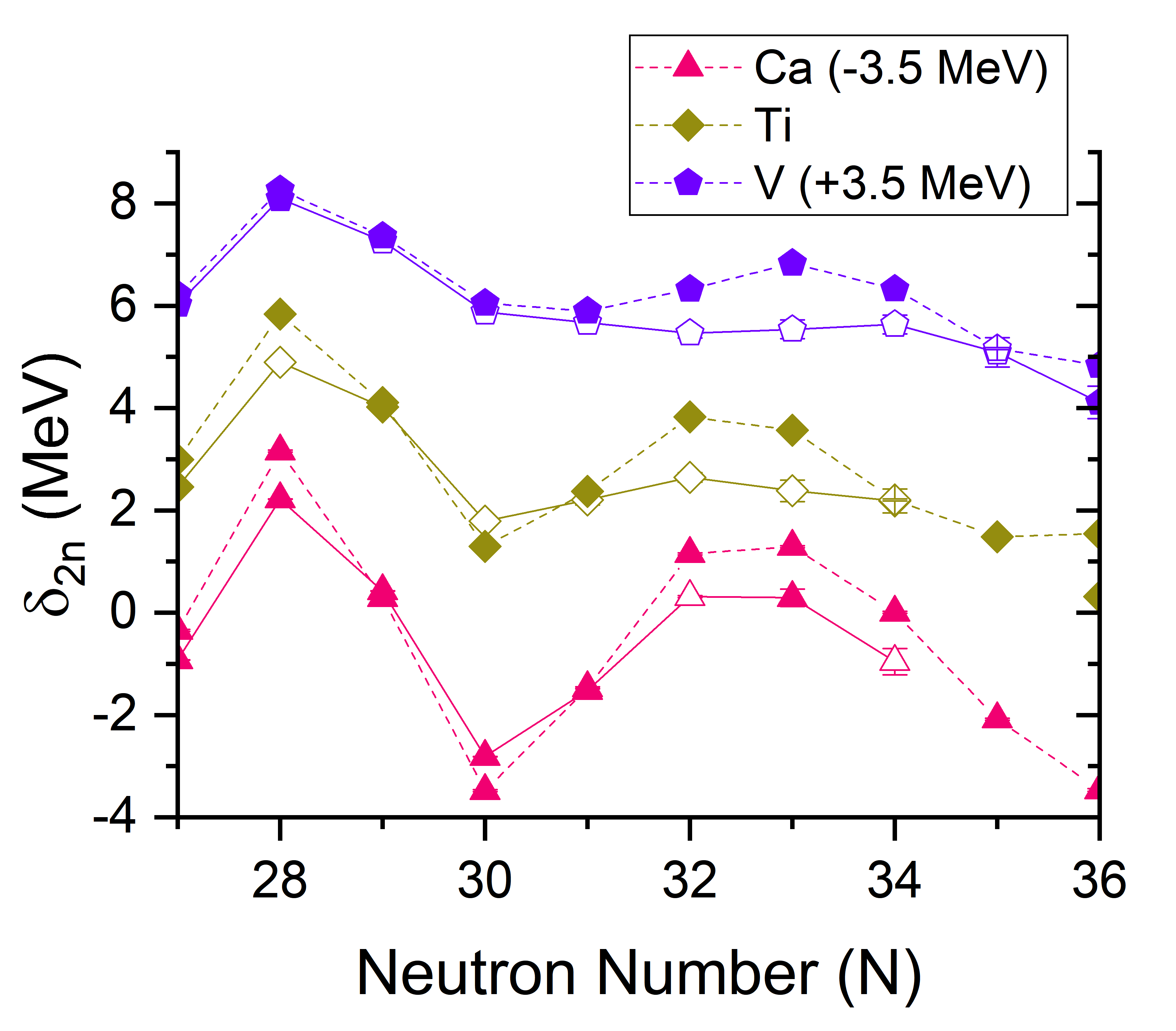}
        \caption{A graph comparing $\delta_{2n}$ experimental values (solid lines) to theoretical calculations (dashed lines) from \cite{Stroberg2021}. Values are offset by amounts as presented in the legend. Unfilled symbols represent values from this work.}
        \label{fig:model_comp}
    \end{center}
\end{figure}

A comparison of our results to recent calculations using the valence-space in-medium similarity renormalization group (VS-IMSRG) \cite{Stroberg2021} is presented in Figure \ref{fig:model_comp}. The VS-IMSRG calculations generally overshoot the experimental results, but follow the overall trends seen in the mass surface, including the disappearance of the $N = 32$ shell closure at higher proton numbers. Additionally, no evidence for a $N = 34$ shell closure is seen in the calculations. More refinement of ab-initio calculations is ultimately needed for a complete picture of the $N = 32$ and $N = 34$ region.

\section{Summary}

Measurements of neutron-rich Ca, Ti and V isotopes were performed at the TITAN facility in Canada using its MR-ToF-MS and the LEBIT facility in the U.S. using its Penning trap mass spectrometer. These results, totaling 13 mass measurements, include eight masses which increase precisions with respect to previous measurements. These measurements refine the nuclear mass surface around $N = 32$ and $N = 34$, and confirm the waning of the $N = 32$ shell closure as $Z = 20$ is exceeded. The refined mass surface also does not support the presence of an $N = 34$ shell closure. Mass data on more neutron-rich nuclides is ultimately needed to further understand the nuclear structure of the region.

\begin{acknowledgments}

We would like to thank J. Lassen and the laser ion source group at TRIUMF for their development of the relevant laser scheme as well as the NSCL staff, the ISAC Beam Delivery group, and M. Good for their technical support. This work was supported by the Natural Sciences and Engineering Research Council (NSERC) of Canada under Grants No. SAPIN-2018-00027, No. RGPAS-2018-522453,
and No. SAPPJ-2018-00028, the National Research Council (NRC) of Canada through TRIUMF, the U.S. National Science Foundation through Grants No. PHY-1565546, No. PHY-2111185, and No. PHY-1811855, the U.S. Department of Energy, Office of Science under Grants No. DE-FG02-93ER40789 and No. DE-SC0015927, the German Research Foundation (DFG), Grant No. SCHE 1969/2-1, the German Federal Ministry for Education and Research (BMBF), Grants No. 05P19RGFN1 and No. 05P21RGFN1, and the Hessian Ministry for Science and Art through the LOEWE Center HICforFAIR, by the JLU and GSI under the JLU-GSI strategic Helmholtz partnership agreement. E.D. acknowledges financial support from the U.K.-Canada Foundation.

\end{acknowledgments}

\bibliography{library}
    
\end{document}